\documentclass[amsmath,amssymb,superscriptaddress,longbibliography,prl,twocolumn,floatfix,citeautoscript]{revtex4-2}

\setcitestyle{super}

\usepackage{graphicx}
\usepackage{xcolor}
\usepackage{comment}
\usepackage[normalem]{ulem}
\usepackage{hyperref}
\hypersetup{
    colorlinks=true,
    urlcolor= blue,
    citecolor=blue,
    linkcolor= blue,
    bookmarksopen=false,
    }

\definecolor{lightgray}{gray}{0.6}
\newif\ifptitle
\newif\ifpnumber
\newcounter{para}
\newcommand\ptitle[1]{\par\refstepcounter{para}
{\ifpnumber{\noindent\textcolor{lightgray}{\textbf{\thepara}}\indent}\fi}
{\ifptitle{\textbf{[{#1}]}}\fi}}
\pnumbertrue  % comment this line to hide paragraph numbers

\newcommand{\heng}{School of Engineering \& Applied Sciences, Harvard University, Cambridge, Massachusetts 02138, USA}
\newcommand{\hphys}{Department of Physics, Harvard University, Cambridge, Massachusetts 02138, USA}

\newcommand{\mpi}{Max Planck Institute for Chemical Physics of Solids, Dresden, Germany}
\newcommand{\wphys}{Department of Physics, University of Warwick}

\begin{document}

\title{Nanoscale variation of the Rashba energy in BiTeI
}

\author{Ruizhe Kang}
\affiliation{\heng}

\author{Jian-Feng Ge}
\altaffiliation[Present address: ]{\mpi}
\affiliation{\hphys}

\author{Yang He}
\affiliation{\hphys}

\author{Zhihuai Zhu}
\affiliation{\hphys}

\author{Daniel T. Larson}
\affiliation{\hphys}

\author{Mohammed Saghir}
\affiliation{\wphys}

\author{Jason D. Hoffman}
\affiliation{\hphys}

\author{Geetha Balakrishnan}
\affiliation{\wphys}

\author{Jennifer E. Hoffman}
\email[]{jhoffman@physics.harvard.edu}
\affiliation{\hphys}
\affiliation{\heng}

\date{\today}

\begin{abstract}
BiTeI is a polar semiconductor with strong spin-orbit coupling (SOC) that produces large Rashba spin splitting. Due to its potential utility in spintronics and magnetoelectrics, it is essential to understand how defects impact the spin transport in this material. Using scanning tunneling microscopy and spectroscopy, we image ring-like charging states of single-atom defects on the iodine surface of BiTeI. We observe nanoscale variations in the Rashba energy around each defect, which we correlate with the local electric field extracted from the bias dependence of each ring radius. Our data demonstrate the local impact of atomic defects on the Rashba effect, which is both a challenge and an opportunity for the development of future nanoscale spintronic devices.
\end{abstract}
% Need to rewrite a little more generally & grandiosely if we're going to submit to Nature Physics.
% https://www.nature.com/documents/nature-summary-paragraph.pdf
% For a bigger sell, don't focus on details like the iodine surface, or the mechanism of how we correlate these things (bias dependence of ring) - but focus more on the result.

\maketitle

\ptitle{SOC is important} Spin-orbit coupling (SOC) plays a crucial role in the rich phenomena observed in condensed matter systems \cite{Manchon_NM2015, Soumyanarayanan_Nat2016, Bihlmayer_NP2022}. The interplay between SOC and broken symmetries gives rise to numerous electronic phases including topological surface states \cite{Hasan_RMP2010} and the quantum spin Hall effect \cite{Kane_PRL2005, Bernevig_PRL2006, Bernevig_Science2006, Konig_Science2007}. When inversion symmetry is broken at surfaces or interfaces, the relativistic interaction between electron spin and momentum splits each energy band into two separate bands with opposite spin polarization \cite{Rashba_SPSS1960, Bychkov_JETP1984}. This phenomenon, known as the Rashba effect, is a starting point to engineer a variety of emergent phenomena driven by SOC. The Rashba Hamiltonian has the form
\begin{equation}
H_{R} = \alpha_{R} \boldsymbol{\hat{z}} \cdot (\boldsymbol{k} \times \boldsymbol{\sigma}),
\label{br}
\end{equation}
where $\alpha_{R}$ is the Rashba parameter that quantifies the strength of the Rashba effect, $\boldsymbol{k}$ and $\boldsymbol{\sigma}$ are the electron momentum and spin, respectively, and $\boldsymbol{\hat{z}}$ is the unit vector perpendicular to the surface. The Hamiltonian in Eq.~(\ref{br}) is of the same form as the Dirac Hamiltonian describing the surface states of topological materials \cite{Hasan_RMP2010}. This leads to a helical spin-polarized Fermi surface, which enables the conversion between charge current and spin current through the direct and inverse Edelstein effects \cite{Edelstein_SSC1990}. The ability to harness and control Rashba effects has tremendous potential for future applications including spin transport devices \cite{Cai_NanoLett2022, Isasa_PRB2016}, spin field-effect transistors \cite{Koo_Science2009, Chuang_NatNano2015}, and spin-orbit torque magnetic random access memory devices \cite{Koo_AdvMat2020}.

\ptitle{Defects \& charging rings} Several challenges must be overcome before Rashba SOC can be employed in practical spintronic devices. Atomic defects can change the electronic properties \cite{Fiedler_NJP2014, Li_npj2020}, tune the chemical potential \cite{hsieh_Nature2009, Beidenkopf_NP2011}, and alter the spin dynamics \cite{Farzaneh_arxiv2018}. Thus, it is essential to understand the effect of single atomic defects on Rashba SOC. Scanning tunneling microscopy and spectroscopy (STM/STS) are powerful tools for this purpose. STM can locate single atom defects, while STS -- differential conductance measurements -- can be used to map their surrounding electron behavior. For example, charge states have been observed around atomic defects on the surface of semiconductors \cite{Teichmann_PRL2008, Wijnheijmer_PRL2009, Tian_PRB2019, Ruan_MTP2020}, Mott insulators \cite{Battiisti_PRB2017} and topological insulators \cite{Song_PRB2012}.

\ptitle{BiTeI has huge Rashba parameter} BiTeI is a non-centrosymmetric semiconductor with a strong Rashba effect both in the bulk and at the surface \cite{Ishizaka_NM2011, Landolt_PRL2012, Kohsaka_PRB2015, Colletta_PRB2014}. Angle-resolved photoemission spectroscopy (ARPES) showed that BiTeI has one of the largest known Rashba parameters, $\alpha_R$ = 3.8 eV\,\AA{} \cite{Ishizaka_NM2011}, which makes it a promising candidate for use in spintronic devices. It was recently proposed that atomic defects in BiTeI could tune the bulk Rashba parameter from 0 to 4.05 eV\,\AA{}  \cite{Li_npj2020}. This theory provides the exciting possibility of using dopants to engineer Rashba materials. However, the local impact of atomic defects on the Rashba effect in BiTeI has not been observed experimentally.

\begin{figure*}
    \includegraphics[width=\textwidth]{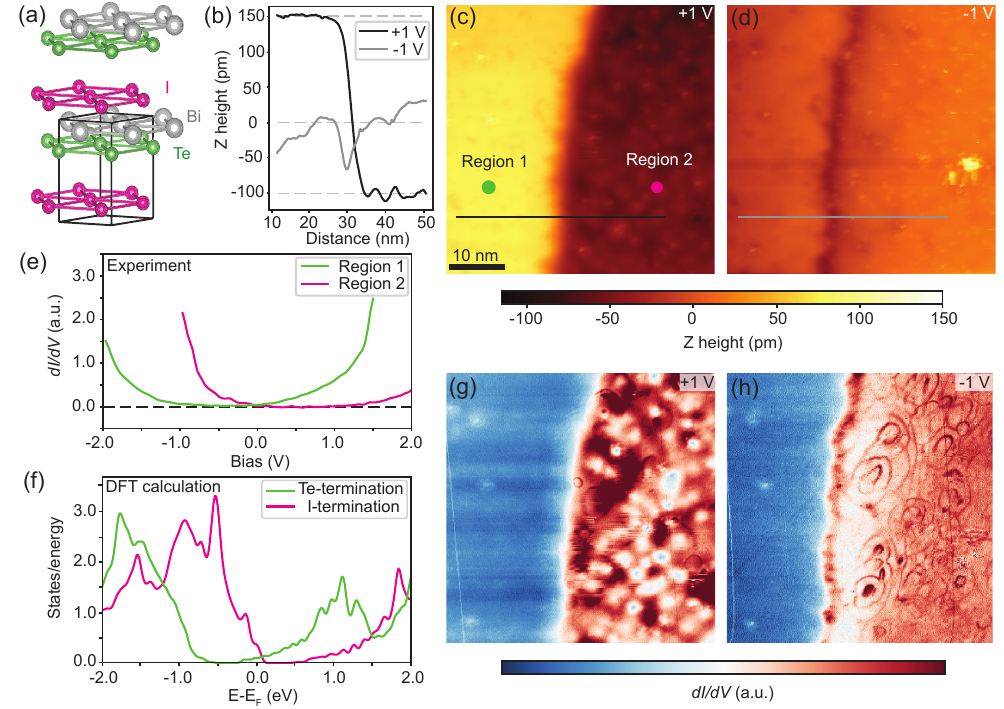}
 	\caption{Topography and spectroscopy on I and Te terminated BiTeI surfaces. (a) Crystal structure of BiTeI. (b) $z$ height trace across the black and grey lines in (c) and (d), respectively. (c) Topography measured in constant current mode around a step edge (sample bias $V_s = +1$ V, current setpoint $I_s = 50$ pA). (d) Same field of view as (c) with opposite sample bias ($V_s = -1$ V, $I_s = 50$ pA).  (e) $dI/dV$ spectra measured on the two BiTeI surface terminations indicated by green and magenta dots in (c). (f) DFT calculation showing the density of states projected onto I and Te surfaces. (g),(h) Differential conductance maps at $V_s = +1$ V and $V_s = -1$ V (lock-in bias modulation $V_{\mathrm{rms}}$ = 20 mV at 1.115 kHz) with the same field of view as (c) and (d).}
	\label{fig:fig1}
\end{figure*}

\ptitle{Here we show ...} Here we use STM and STS to characterize the electronic behavior around single atom defects in BiTeI. First, we observe the charge states of the defects using the tip-induced band bending (TIBB) effect, which manifests as circular rings surrounding defects on the I-terminated surface. We use the bias-dependence of the ring size as a local probe of the electric field at the surface. Second, the Rashba band splitting leads to van Hove singularities whose measured spectral width gives the local Rashba energy. We discover that the local Rashba energy is correlated with the local electric field, demonstrating a pathway for nanoscale Rashba engineering via atomic dopant control.
 
 \begin{figure*}
	\includegraphics[width=\textwidth]{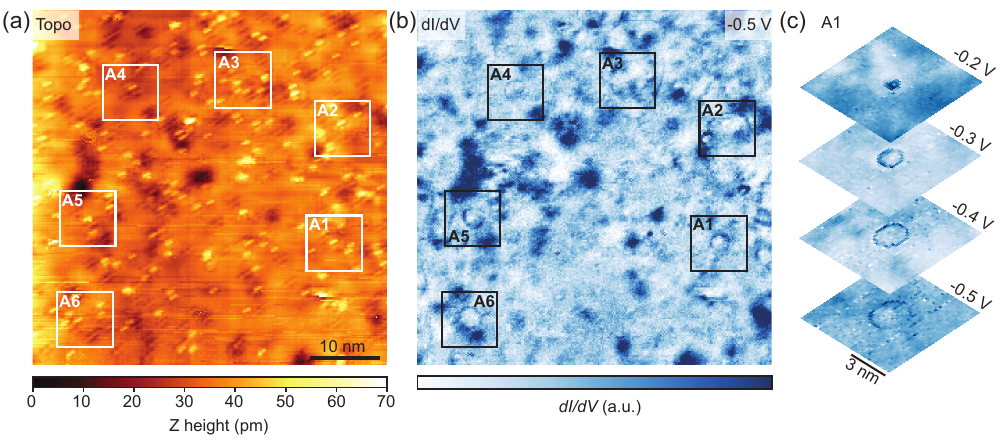}
 	\caption{Tip-induced band bending effect on I-terminated BiTeI surface. (a) Topography ($V_s = -0.5$ V, $I_s = 150$ pA) and (b) simultaneously-acquired $dI/dV$ map ($V_{\mathrm{rms}} = 20$ mV) on the I-terminated surface. The white and black squares highlight the regions with charging rings. (c) Bias-dependent behavior for the charging ring in region A1, showing an increase in the ring radius with decreasing bias voltage.}
	\label{fig:fig2}
 \end{figure*}

\ptitle{BiTeI has two terminations} Single crystals of BiTeI were grown by the Bridgman method \cite{Ishizaka_NM2011}, cleaved in cryogenic ultra-high vacuum conditions, and immediately inserted into a home-built STM for measurement at 4.6 K using a mechanically cut Pt-Ir tip. BiTeI is composed of single layers of Bi atoms in a hexagonal lattice, covalently bonded to hexagonal layers of Te and I above and below, as shown in Fig.~\ref{fig:fig1}(a). BiTeI naturally cleaves between the Te and I layers due to the weak bonding between those two layers.
This leads to two types of cleaved surfaces, terminated by I or Te. To identify the surface termination of our sample, we acquire topographic maps around a step edge at two different sample biases, $V_s = +1$ V in Fig.~\ref{fig:fig1}(c) and $V_s = -1$ V in Fig.~\ref{fig:fig1}(d).
The strong bias-dependence of the apparent step height shown in the linecuts of Fig.\ \ref{fig:fig1}(b) indicates that the terraces on either side of the step are chemically different. 
Differential conductance spectra ($dI/dV$) on both terraces in Fig.~\ref{fig:fig1}(c) confirm this argument. Figure \ref{fig:fig1}(e) shows that region 2 is gapped at positive bias (above the Fermi level $E_F$), while region 1 is gapped at negative bias (below $E_F$). Density functional theory (DFT) calculations of the density of states projected onto these two terminations, in Fig.~\ref{fig:fig1}(f), show that the I-termination is gapped for positive energies while the Te-termination is gapped for negative energies. Thus we conclude that region 2 is I-terminated, while region 1 is Te-terminated.
 
\ptitle{dI/dV maps} We acquire $dI/dV$ maps at both positive bias [Fig.~\ref{fig:fig1}(g)] and negative bias [Fig.~\ref{fig:fig1}(h)] in the same field of view. For $V_s = -1$ V there is high spectral intensity along the step edge due to the ambipolarity of the sample surface. The I-termination acts as a \emph{p}-type semiconductor while the Te-termination is \emph{n}-type, so together they form a nanoscale \emph{p-n} junction. The spectral intensity we observe between the two regions originates from the depletion layer formed at the \emph{p-n} junction interface, in agreement with previous spectroscopic measurements at BiTeI step edges \cite{Colletta_PRB2014}. 

\ptitle{6 charging rings} Here we focus on ring-like structures around defects at negative bias on the I-terminated surface due to TIBB \cite{Teichmann_PRL2008, SI} shown in Fig.~\ref{fig:fig1}(h). Figures \ref{fig:fig2}(a) and (b) show the simultaneously-acquired topography and $dI/dV$ maps of the I-terminated surface. Within the 50 $\times$ 50 nm$^2$ scan region, six charging rings are visible, labeled A1 to A6. The radius of the charging rings increases as the sample bias decreases, as shown in Fig.~\ref{fig:fig2}(c) for the ring in region A1.

  \begin{figure*}
  	\includegraphics[width=\textwidth]{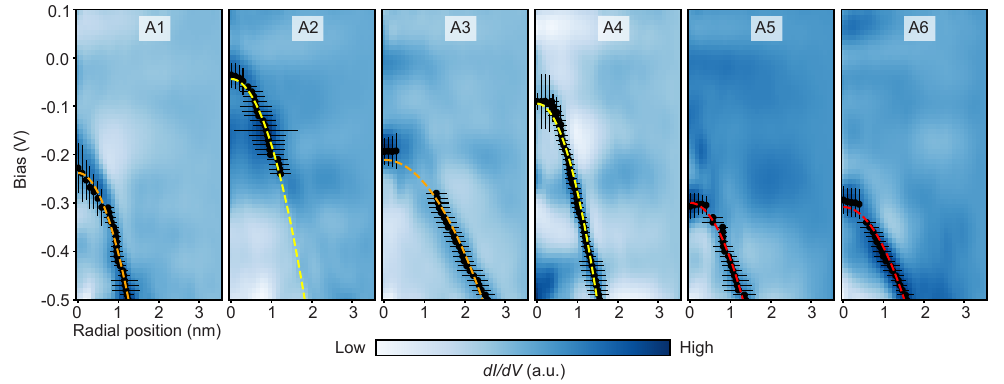}
  	\caption{Bias-dependent charging ring radius. The data points are extracted at the peak position from vertical or horizontal line cuts and the error bars are determined by the difference between the peak and its following valley position (see Supplementary Information\cite{SI} for detailed data processing). Dashed curves indicate the quadratic fits to the full data set for each region.}
 	\label{fig:fig3}
  \end{figure*}

\ptitle{Justify the quadratic fitting} In Fig.~\ref{fig:fig3} we plot the azimuthally-averaged differential conductance as a function of distance from the ring center for all six regions indicated in Fig.~\ref{fig:fig2}(b). For small radii, we can approximate the sample bias dependence of ring radius with a quadratic function\cite{Teichmann_PRL2008, Feenstra_VST2003}, 
\begin{equation}
V_s = ar^2 + V_0
\label{quadrafit}
\end{equation}
where $V_s$ is the sample bias, and $r$ is the ring radius. The two fitting parameters are $a$, which encodes the $E$-field dependence of the surface electrons, and $V_0$, which encodes the ionization energy of the defect.
The quadratic fits are plotted as dashed lines in Fig.~\ref{fig:fig3}.

\ptitle{Intercept describes the defect type, slope represents effective electric field} The defects fall into three groups: A5 and A6 ($V_0 \sim -0.3$ V), A1 and A3 ($V_0 \sim -0.2$ V), and A2 and A4 ($V_0 \sim -0.1$ V).
However, the leading coefficients differ, even for defects with similar intercepts. Though the absolute values of the intercepts and the leading coefficients depend sensitively on tip-sample distance \cite{Ruan_MTP2020}, the tip-sample distance is kept constant during the $dI/dV$ map, which makes it possible to qualitatively compare the difference.
We propose that the different intercepts represent different surface potentials while the leading coefficients are local probes for the total electric field.

\begin{figure}[t!]
	\includegraphics[width=\columnwidth]{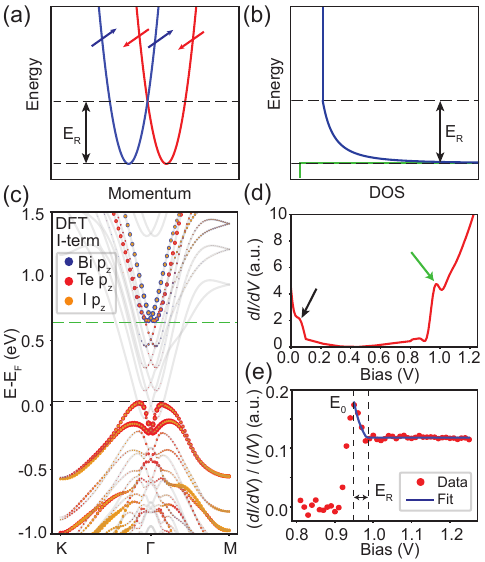}
 	\caption{van Hove singularity caused by Rashba splitting. (a,b) Schematic representation of the band splitting caused by the Rashba effect. The blue and red parabola in (a) represent opposite spin polarization. (b) shows the corresponding density of states of the band structure in (a), which has a van Hove singularity at the energy of the band bottom. (c) DFT calculation of the electronic structure of a BiTeI slab showing projections onto the $p_z$ orbitals of the I-terminated surface. The zero of energy is the Fermi level of the entire slab. The black and green dashed lines indicate the band top and bottom, respectively, where the van Hove singularities should appear. (d) $dI/dV$ spectrum measured on the I-terminated surface above the Fermi energy ($I_s$= 150 pA, $V_s$= -500 mV, $V_{\mathrm{rms}} = 20$ mV). Black and green arrows indicate the van Hove singularities. (e) Normalized $dI/dV$ spectrum averaged over 3.3 $\times$ 3.3 nm$^2$ around the defect in A1 region. The blue fitted curve uses Eq.~(\ref{dos})}
	\label{fig:fig4}
 \end{figure}

\ptitle{Rashba effect results in van Hove singularity} Rashba first proposed a model describing a 2D free electron gas (2DEG) in a perpendicular electric field \cite{Rashba_SPSS1960}. The parabolic band of the 2DEG splits into two bands with opposite spin polarization. The resulting van Hove singularity at the band bottom is shown schematically in Fig.~\ref{fig:fig4}(a,b). The density of states in this model is given by:
\begin{equation}
D(E) = 
\begin{cases}
      \displaystyle \frac{\displaystyle m^*}{\displaystyle \pi \displaystyle \hbar^2}=\mathrm{\displaystyle const} & E>E_0+E_R\\
      \\
      \displaystyle \frac{\displaystyle m^*}{\displaystyle \pi \displaystyle \hbar^2} \sqrt{\frac{\displaystyle E_R}{\displaystyle E-E_0}} & E_0<E<E_0+E_R\\
    \end{cases}   
    \label{dos}
\end{equation}
where $m^*$ is the effective mass, $E_0$ is the energy at the band bottom, and the Rashba energy $E_R$ defines the energy difference between the crossing point of the bands and the band bottom, which is proportional to $\alpha_R^2$ (see Supplementary Information\cite{SI} for details). This simple model captures the main picture of Rashba physics, though local disorder can add more complexity to the problem\cite{Glazov_PhysicaE2010}. Figure~\ref{fig:fig4}(d) shows a typical $dI/dV$ spectrum on the I-terminated surface. The two van Hove singularities around 50 mV and 950 mV are indicated by arrows. Comparing with DFT calculations of the band structure projected onto the I-terminated surface, shown in Fig.~\ref{fig:fig4}(c), we see that the van Hove singularity at the top of the valence band is composed of Te and I $p_z$ orbitals while the van Hove singularity at the bottom of the conduction band is composed of $p_z$ orbitals from all three elements.

\ptitle{$E_R$ correlates with $a$} Though the splitting of the surface state is not directly observable using STM \cite{Petersen_SurfSci2000}, the local Rashba energy can be extracted by fitting $dI/dV$ at the surface \cite{Ast_PRB2007, Ast_JESRP2015}. To investigate how the Rashba energy varies between different defect sites, we fit the van Hove singularities peaked at the conduction band bottom with the density of states functions in Eq.~(\ref{dos}). All spectra are averaged over a 3.3 $\times$ 3.3 nm$^2$ region centered within the charging rings. We first normalize the data by dividing the differential conductance ($dI/dV$) by its conductance ($I/V$), then subtract a linear background. Figure \ref{fig:fig4}(e) shows the fit for the A1 region, where the extracted Rashba splitting $E_R$ is 57.9 $\pm$ 10 meV, which is smaller than the bulk value of 100 meV \cite{Ishizaka_NM2011}. We carry out the same fitting procedure for all six charging ring regions identified in Fig.~\ref{fig:fig2} (see Supplementary Information \cite{SI} for details). We observe that the Rashba energy varies among the defects from 34.5 meV (A3) to 77.9 meV (A4).  
Similar spatial fluctuation of the Rashba energy have also been observed on the $p$-type InSb (110) surface, resulting from the bulk acceptor disorder \cite{Bindel_NP2016}.

\begin{figure}[t!]
	\includegraphics[width=\columnwidth]{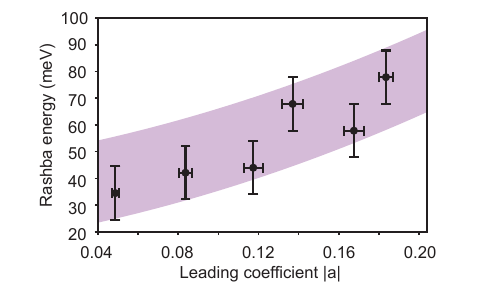}
 	\caption{Correlation between the local Rashba energy and the leading coefficient of the charging ring dispersion. Vertical error bars are energy resolution used in measuring the STS. Horizontal error bars are the standard errors of leading coefficient $a$ from the covariance matrix of the optimally fitted parameters using Eq.~(\ref{quadrafit}). The shaded region is added as a guide to the eye.}
	\label{fig:fig5}
 \end{figure}

\ptitle{Discussion} Intriguingly, we find that the Rashba energies around the defects correlate with the leading coefficient $a$ of the charging ring dispersion, as shown in Fig.~\ref{fig:fig5}. Namely, the larger the leading coefficient of the parabola (i.e., the smaller the opening angle), the larger the local Rashba energy and Rashba parameter. We can understand this behavior qualitatively by considering the Rashba effect on a 2DEG. In this case, the Rashba parameter is proportional to the electric field experienced by the electrons and the Rashba energy is proportional to the electric field intensity squared (see Supplementary Information \cite{SI}).

\ptitle{Efield origin} The electric field results from three intrinsic factors in BiTeI. First, BiTeI has a polar surface with a net electric field from the (BiTe)$^+$ and I$^-$ dipole moment \cite{Butler_NC2014}. Second, surface defects, such as atomic vacancies and adatoms, lead to either accumulation or depletion of charge on the surface, which also contribute to the local electric field. Third, surface electrons affect the electric field through Coulomb screening. All three factors, together with the gating effect from the tip, form a net local electric field. The charging ring states inherently capture the behavior of the surface electrons in response to this local field: the larger the leading coefficient $a$ of the ring dispersion, the larger the effective electric field, which increases the asymmetry of the electron wavefunction near the nucleus. Increased asymmetry enhances the Rashba splitting and Rashba energy \cite{Bihlmayer_SurfSci2006}. 

\ptitle{Conclusion} In conclusion, we observe the charging states of defects on the iodine termination of BiTeI and use the charging ring radius as a probe of the local electric field. We extract the local Rashba energies from $dI/dV$ spectra by fitting the van Hove singularities originating from the Rashba band splitting. We demonstrate that the local Rashba energies correlate with the effective electric field. Our work shows that atomic dopants can modify the local Rashba energy in BiTeI, which presents an exciting opportunity for engineering nanoscale spintronic devices.

\section*{Acknowledgements}
We thank Aaron Coe, Evgeny Ya.\ Sherman, Eslam Khalaf, and Shiang Fang for insightful discussions. R.K.\ is funded by the Department of Energy through Award No.\ DE-SC0020128.
D.T.L.\ acknowledges funding from the STC Center for Integrated Quantum Materials, NSF Grant No.\ DMR-1231319 and NSF Award No.\ DMR-1922172.
DFT calculations were performed on the FASRC Cannon cluster supported by the FAS Division of Science Research Computing Group at Harvard University. The work at the University of Warwick was supported by EPSRC, UK through Grants EP/T005963/1 and EP/L014963/1.

\bibliography{refs}

\end{document}

\title{Supplementary material for Nanoscale variation of the Rashba energy in BiTeI}

\author{Ruizhe Kang}
\affiliation{\heng}

\author{Jian-Feng Ge}
\altaffiliation[Present address: ]{\mpi}
\affiliation{\hphys}

\author{Yang He}
\affiliation{\hphys}

\author{Zhihuai Zhu}
\affiliation{\hphys}

\author{Daniel T. Larson}
\affiliation{\hphys}

\author{Mohammed Saghir}
\affiliation{\wphys}

\author{Jason D. Hoffman}
\affiliation{\hphys}

\author{Geetha Balakrishnan}
\affiliation{\wphys}

\author{Jennifer E. Hoffman}
\email[]{jhoffman@physics.harvard.edu}
\affiliation{\hphys}
\affiliation{\heng}

\date{\today}
\maketitle

\section{Appendix A: Atomically Resolved Defects}
% on the I-terminated surface of BiTeI

\begin{figure*}[!htb]
    \includegraphics[width=16cm]{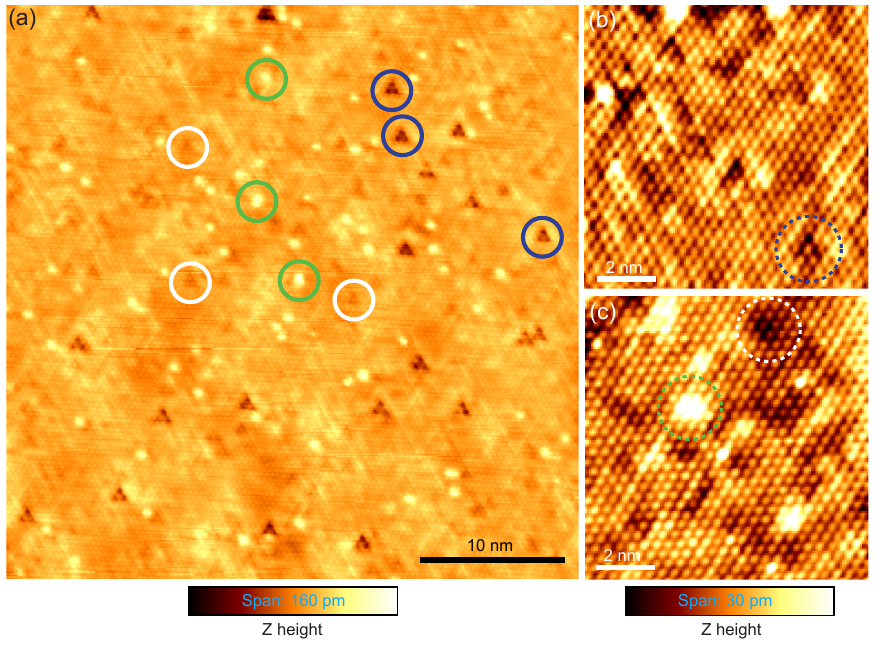}
 	\caption{Atomically-resolved topography on the I termination of BiTeI (a) Several defects are visible on the I-terminated surface. Examples are surface atomic vacancies (blue circles), surface adatoms (green circles) and subsurface defects (white circles). (Sample bias $V_s = 350$ mV, current setpoint $I_s = 40$ pA.) (b), (c) Zoomed-in scans for defects on I-terminated surface. ($V_s = 400$ mV, $I_s = 50$ pA.) (b) and (c) share the same colorscale.}
	\label{fig:figs1}
\end{figure*}

\ptitle{defect types in topography} Atomically resolved images in Fig.~\ref{fig:figs1} show intrinsic defects on the I-terminated surface of BiTeI. We identify three different types of defects, which may correspond to the three values of charging ring energy offset $V_0$ shown in main text Fig.~3.

\section{Appendix B: Tip-Induced Band Bending}

\begin{figure*}[!htb]
    \includegraphics[width=16cm]{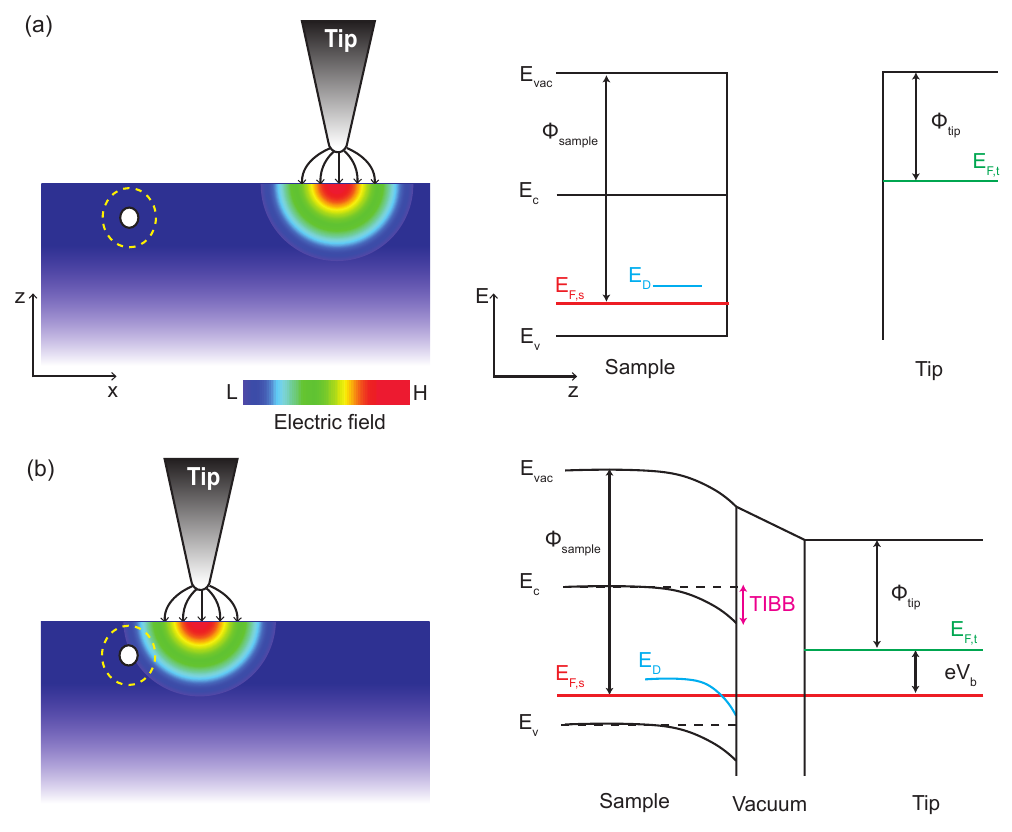}
 	\caption{Schematic of the tip-induced band bending (TIBB) effect on a semiconductor surface. (a) When the tip is far away from the defect and (b) when the tip is close to the defect. E$_c$ and E$_v$ are conduction band bottom energy and valance band top energy, respectively. $\Phi_{sample}$ and $\Phi_{tip}$ are the work function of sample and tip. E$_{vac}$ is the vacuum energy, and E$_D$ is the energy level of the defect. E$_{F,s}$ and E$_{F,t}$ are the Fermi energy of the sample and the tip, respectively. V$_b$ is the bias applied to the sample.}
	\label{fig:figs2}
\end{figure*}

\ptitle{Tip-induced band bending} Tip-induced band bending (TIBB) arises from poor charge screening on the sample surface. The mechanism of TIBB is shown schematically in Fig.~\ref{fig:figs2}. On surfaces with low charge carrier density, such as semiconductors and topological insulators, TIBB manifests as rings around defects in the differential conductance ($dI/dV$) map. Defects on the semiconductor surfaces act as either electron donors or acceptors, which form energy levels within the gap. Since a typical tip-sample distance for STM is a few angstroms, the electric field between the tip and sample has a gating effect on the sample surface. The electric potential gradient across the tunneling barrier includes contributions from both the applied bias and work function difference between the sample and the tip. When the tip is far from the defect, the defect level and the local band structure are not affected by the tip [Fig. \ref{fig:figs2}(a)]. Previous literature has reported the work function is about 6 eV on the I-terminated surface\cite{Kohsaka_PRB2015}, while the typical work function for a Pt-Ir tip is 5.2 eV and we apply a bias of -0.5 V to the sample. Then, as the tip approaches the defect, the gating effect from the tip bends the local bands down. Once the defect level is bent down below the sample Fermi surface, there will be a abrupt change in the number of states that the tip can tunnel into [Fig. \ref{fig:figs2}(b)]. Thus, there will be a change of the tunneling current, when the tip is at the critical lateral distance from the defect. This process is independent of the direction, thus we should observe a ring in the differential conductance map. Similar TIBB-induced charging rings have previously been observed in a variety of systems, including the \{110\} surface of silicon doped GaAs\cite{Teichmann_PRL2008, Wijnheijmer_PRL2009}, potassium adatoms on the surface of black phosphorus\cite{Tian_PRB2019}, and charged defects on surface of WS$_2$\cite{Ruan_MTP2020}.

\newpage

\section{Appendix C: \texorpdfstring{$dI/dV$}{dI/dV} Data Acquisition \& Analysis}

\ptitle{$dI/dV$ data acquisition} The topography and $dI/dV$ map shown in Fig.~2 were measured simultaneously using a feedback loop with sample bias $V_s = -0.5$ V and current setpoint $I_s = 150$ pA. After fixing the tip height at each pixel, the feedback loop was turned off, and $dI/dV$ was measured using a lock-in technique with a bias modulation $V_{\mathrm{rms}} = 20$ mV at frequency 1.115 kHz. The resulting $dI/dV$ map in Fig.~2(b) was smoothed using a mean filter that set each pixel equal to the average value of its 4 nearest neighbors if the pixel is not within $\pm$6$\%$ of the averaged value in order to reduce the salt-and-pepper noise.

\begin{figure*}[!htb]
    \includegraphics[width=16cm]{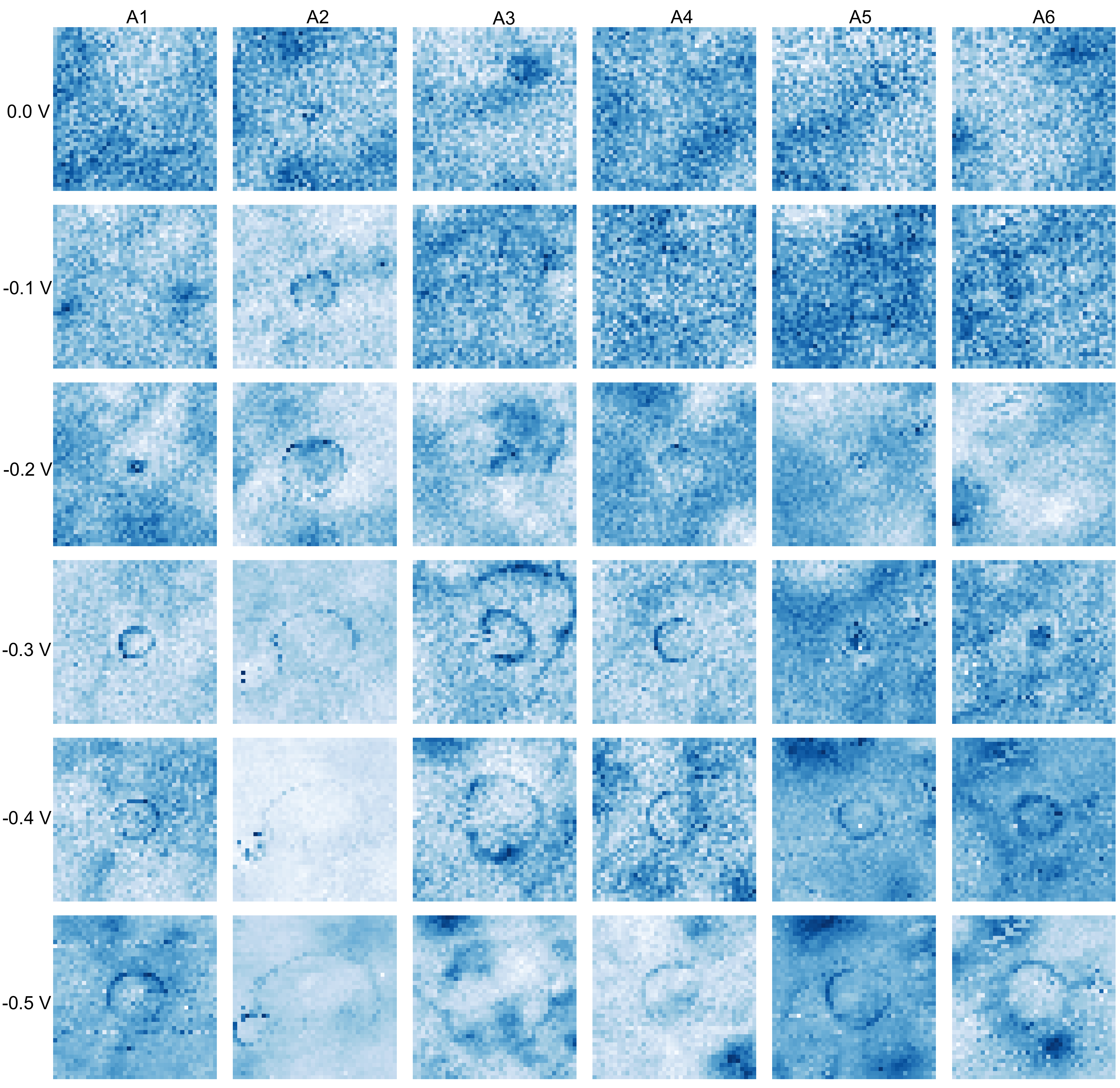}
 	\caption{Raw data of $dI/dV$ maps for 6 charging ring regions at 6 sample biases. $V_s = -0.5$ V; $I_s = 150$ pA, $V_{\mathrm{rms}} = 20$ mV.}
	\label{fig:figs3}
\end{figure*}

\newpage

\ptitle{Charging ring analysis} In Fig.~3 and Fig.~S4, each energy layer of the resulting $dI/dV$ map was smoothed using a low pass filter to increase the signal-to-noise ratio. Then the maps were azimuthally-averaged and gradient filtered to help extract 1D ridge dispersion from the 2D image plot\cite{He_RSI2017}. The data points were extracted from the first local maximum peak followed by a local minimum of either vertical or horizontal linecuts of 1D data. The error bars were determined by 40$\%$ of the difference of local maxima and minima position. 

\begin{figure*}[!htb]
    \includegraphics[width=\textwidth]{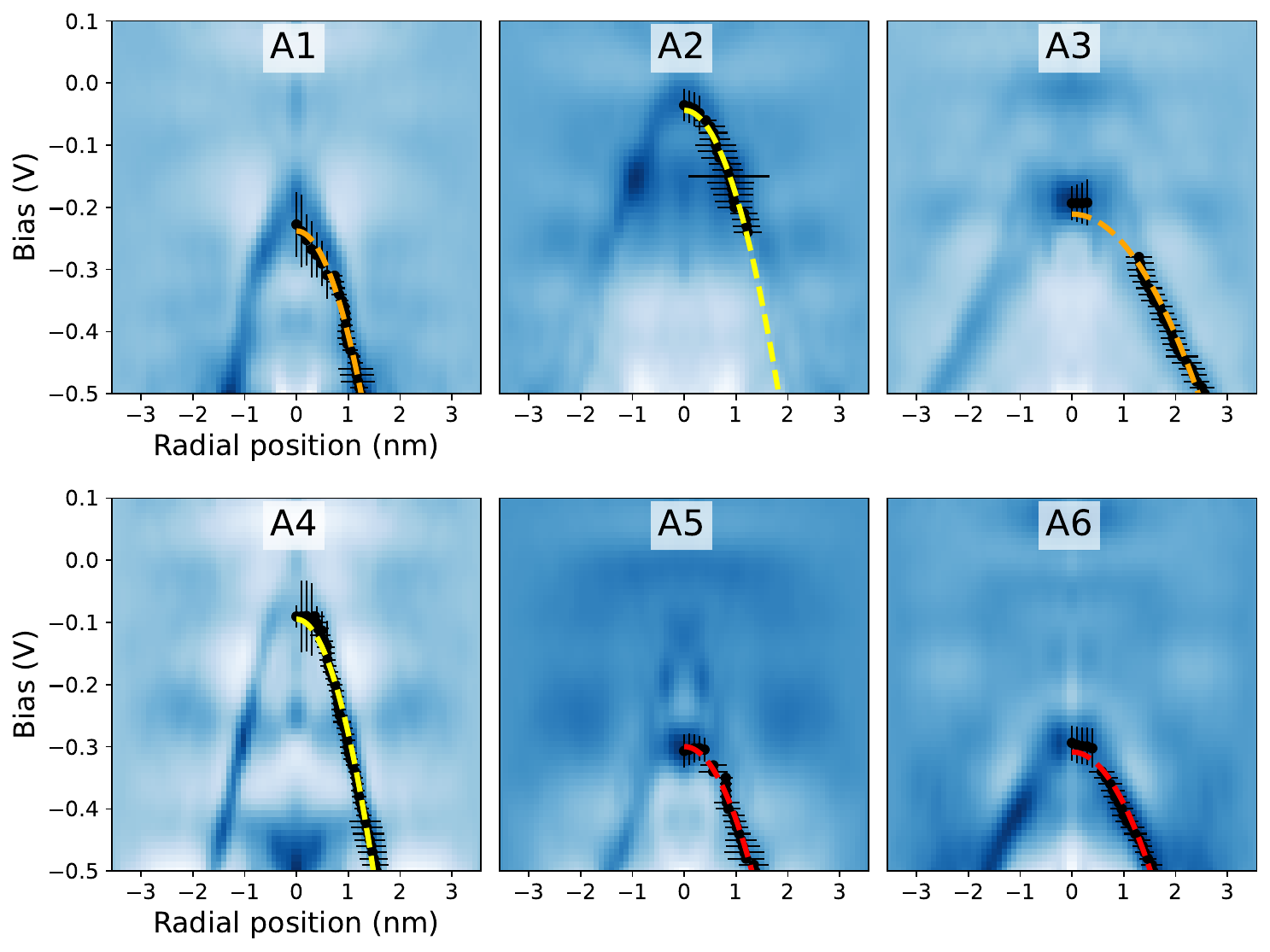}
 	\caption{Bias-dependent charging ring radius. In each plot, the right part is the same as is shown in Fig.~3 in the main text with data points and the fitted curves. The left half shows the unobscured data, mirrored along the axis. Note that a second charging ring is observed in region A3.}
	\label{fig:figs4}
\end{figure*}

\newpage
%\section{Appendix E: Data processing}

\ptitle{Fitting $dI/dV$ spectra} The spectra for all 6 regions are averaged over a 3.3 $\times$ 3.3 nm$^2$ region centered at the center of the charging rings. All spectra are zero-shifted to avoid negative differential conductance. We normalized the data by dividing $dI/dV$ by $I/V$ to make the peak more prominent. We fitted and subtracted a linear background to the data above the band crossing point. We fit the Rashba energy $E_R$ and the band bottom energy $E_0$ using Eq.~(3) in the main text. We constrained the fit value of $E_0$ to be between 920 meV and 950 meV.

\renewcommand\arraystretch{1.8}
\begin{table}[htb]
\begin{ruledtabular}
    \centering
    \caption{
    \label{table:1}
    Fitted parameters of 6 charging rings}
    
    \begin{tabular}{c|cccccc}
                                    & \textbf{A1}   & \textbf{A2}   &   \textbf{A3} & \textbf{A4}   & \textbf{A5}   & \textbf{A6}   \\
\hline
Rashba energy $E_R$ (meV)           & 57.9          & 67.9          & 34.5          & 77.9          & 44.0          & 42.1          \\
Band bottom energy $E_0$ (meV)           & 925          & 923          & 942          & 920          & 946          & 920         \\
Parabola leading coefficient $|a|$  & 0.168         & 0.137         & 0.049         & 0.184         & 0.117         & 0.084         \\
Parabola intercept $V_0$ (V)          & -0.238        & -0.044        & -0.211        & -0.095        & -0.300        & -0.308               
    \end{tabular}
\end{ruledtabular}
\end{table}

\begin{figure*}[!htb]
    \includegraphics[width=\textwidth]{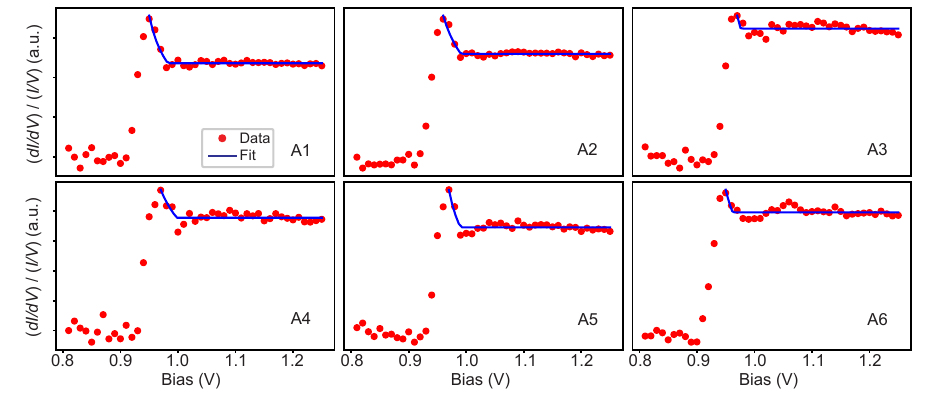}
 	\caption{Fitting for the van Hove singularities for all 6 regions shown in the main text.}
	\label{fig:figs5}
\end{figure*}

\section{Appendix D: Computational Details}

DFT calculations including spin-orbit coupling were performed with the \texttt{VASP} package\cite{Kresse1996efficient,Kresse1996efficiency} using the Perdew-Burke-Ernzerhof exchange-correlation functional\cite{Perdew1996gga}, the DFT-D3 method (zero damping) to account for van der Waals forces between layers\cite{Grimme2010consistent}, and projector augmented wave (PAW) pseudopotentials with 15, 6, and 7 valences electrons for Bi, Te, and I, respectively\cite{Blochl1994paw,Kresse1999paw}.
The crystal surface was modeled with a slab structure with 30~\AA{} of vacuum space and an energy cutoff of 300 eV.

The BiTeI crystal structure can be visualized as a Bi layer sandwiched between a layer each of Te and I forming a triple-layer (TL) unit. The curves in Fig.~1(f) were derived from a slab with 9 TLs by projecting the total DOS onto the atoms in the single TL at one surface of the slab, one with I termination and the other with Te termination. Momentum space was sampled with a 30$\times$30$\times$1 $\Gamma$-centered grid.
%.../projects/BiTeI/9TL/top-bot-relax/dense-kpts/DOSCAR
The band structure in Fig.~4(c) was calculated for a slab with 5 TLs, and the orbital character determined by projecting the charge density onto the $p_z$ orbitals of the atoms in TL of the I-terminated surface. The self-consistent charge density was determined using a 15$\times$15$\times$1 $\Gamma$-centered $k$-point grid.
%.../projects/BiTeI/5TL/ -5, -4
%Used dipole correction?  No

\section{Appendix E: Toy model for Rashba effect}

Consider a 2DEG with an external electrical field $\vecEz$ applied normally to the 2DEG plane. The Hamiltonian will thus have an extra term that breaks inversion symmetry,
\begin{equation}
H = -\vecEz ez.
\label{1}
\end{equation}
Considering relativistic corrections, the electron moving at velocity $\vec{v}$ in electric field $\Ez$ will feel a magnetic field,
\begin{equation}
\vec{B} = -\frac{1}{c^2}(\vec{v}\times\vecEz).
\label{2}
\end{equation}
Using $\vec{\mu}=-(g \mu_B /2)\vec{\sigma}$ and $\vec{v}=(\hbar/m^*)\vec{k}$, the Rashba spin-orbit coupling Hamiltonian can be written as
% \red{(JEH added $c^2$ in denominator! Please double-check the rest of math with $k_0$ and $E_0$ and $E_R$.)}
%
\begin{equation}
H_{R} = -\vec{\mu}\cdot\vec{B} = \frac{g\mu_{B}\hbar}{2m^*c^2}\vec{\sigma}\cdot(\vecEz\times\vec{k}) = \lambda \Ez \boldsymbol{\hat{z}} \cdot( \vec{k}\times \vec{\sigma}),
\label{4}
\end{equation}
where $\lambda = g\mu_{B}\hbar/(2m^*c^2)$ is the SOC coupling constant, $g$ is the electron g-factor, $\mu_{B}$ is the Bohr magneton, and $m^*$ is the effective mass.
Compared to Eq.~(1) in the main text, we obtain $\alpha_{R} = \lambda\Ez$.
The energy dispersion is
\begin{equation}
E = \frac{\hbar^2}{2m^*}(\vec{k} + \vec{k}_0)^2 + E_0,
\label{5}
\end{equation}
where $k_0 = m^*\alpha_{R}/\hbar^2$ is the momentum shift of each parabolic band away from the $\Gamma$ point, and $E_0$ is the energy offset.
The Rashba energy $E_R$ is given by
\begin{equation}
E_R = \frac{\hbar^2}{2m^*}k_0^2 = \frac{m^*\alpha_R^2}{2\hbar^2} = \frac{m^*\lambda^2}{2\hbar^2} \Ez^2.
\label{6}
\end{equation}
Thus, we can conclude that the Rashba energy is proportional to the electrical field intensity squared ($E_R \propto |\Ez|^2$).

\bibliography{ref_supp}